\documentclass[aps,prb,twocolumn,floatfix,showpacs,showkeys,amssymb]{revtex4}
\usepackage{indentfirst}
\usepackage{amsmath}
\usepackage{graphicx}

\usepackage{amsfonts}

\usepackage{amssymb}

\begin{document}

%\begin{widetext}
%\begin{center}
\title{ { {Photon-assistant Fano resonance in coupled multiple
quantum dots }}}

\author{Wanyuan Xie$^{1}$, Hui Pan$^{1,2}$, Weidong Chu$^{1}$, Wei Zhang$^{1}$
and Suqing Duan$^{1}$
{\footnote{Corresponding author, Email-address:
duan$\_$suqing@iapcm.ac.cn}}}

\affiliation{$^{1}$ Institute of Applied Physics and Computational
Mathematics, P.O. Box 8009, Beijing 100088, China\\
$^{2}$ Department of Physics, Beijing University of Aeronautics
and Astronautics, Beijing 100083, China}
%\end{center}

\begin{abstract}
%\begin{widetext}
\smallskip
Based on calculations of the electronic structure of coupled
multiple quantum dots, we study systemically the transport
properties of the system driven by an ac electric field. We find
{\bf qualitative} difference between transport properties of double
coupled quantum dots (DQDs) and triple quantum dots. For both
symmetrical and asymmetrical configurations of coupled DQDs, the
field can induce the photon-assisted Fano resonances in current-AC
frequency curve in parallel DQDs, and a symmetric resonance in
serial DQDs. For serially coupled triple quantum dots(STQDs), it is
found that the $\Lambda$-type energy level has remarkable impact on
the transport properties. For an asymmetric (between left and right
dots) configuration, there is a symmetric peak due to resonant
photon induced mixing between left/right dot and middle dot. In the
symmetric configuration, a Fano asymmetric line shape appears with
the help of ``trapping dark state". Here the interesting coherent
trapping phenomena, which usual appear in quantum optics, play an
essential role in quantum electronic transport. We provide a clear
physics picture for the Fano resonance and convenient ways to tune
the Fano effects.
\end{abstract}

%\ \ \ \ {\bf PACS: 73.63.Kv, 05.60.Gg, 25.70.Ef}
\maketitle \setcounter{section}{0} \setcounter{equation}{0}
\setcounter{subsection}{0} \setcounter{subsubsection}{0}
%\end{widetext}

\section{Introduction}
The electronic transport in quantum dot structures is of great
interest and has been the subject of active research in recent
years. The strong spacial confinement leads to the discrete energy
spectrum. These quantum dots (so called ``artificial atoms"
)/mutiple coupled quantum dots (MCQDs, so called ``artificial
molecules" ) are very important in understanding fundamental quantum
phenomena. Many interesting phenomena which appear in real atoms and
molecules can be well manifested in these nanostructures. Tunability
provides much more opportunities for exploring richer physics which
is hard to access in real atoms/molecules. It also leads to various
applications. For instance, novel quantum logical gates and
elementary qubits in quantum computers\cite{8,9,10} could be
realized based on these systems.

In these nanostructures, quantum coherence and interference is one
of the most important issues and leads to a lot of interesting
phenomena such as Aharonov-Bohm (AB) oscillations\cite{heib}, Kondo
effect\cite{2,3}, coherent trapping\cite{4,23}, Fano
resonance\cite{5,6,7} and so on. In recent years, double and triple
quantum dots have attracted much attention due to their rich
electric structure and physical phenomena.  An AB interferometer
containing two QDs has been
realized\cite{Holleitner1,Holleitner2,Holleitner3,Chen}. Some
experimental groups have been able to fabricate triple quantum dots
with hight quality \cite{exp1,exp2}, and many relevant theoretical
studies have been conducted on these
systems\cite{11,12,13,14,15,16,Konig,Claro,Kang,Bai,Dong,Lopez}.
%The triple quantum dot is one of the nanostructures on which plenty
%of research has been done. In serially coupled triple quantum dots,
%people have studied the spin-entangled current[11], Kondo
%effect[12-13], Fermi-liquid versus Non-Fermi liquid behavior[14]. In
%parallel coupled triple quantum dots, spectral properties and
%quantum phase transitions[15], Fano resonance and bound states[16]
%have been studied .
In spite of many studies on the multiple quantum dots, relatively
little attention has been paid to the photon-assistant transport in
multiple quantum dots driven by a time-period electric field.
Moreover, periodic driven triple quantum dots with the relevant
three-level structure may have some interesting interference
phenomena, such as coherent trapping, like those in quantum optics.
In this paper we study the photon-assistant transport in MCQDs (see
Fig.1) paying special attention to the consequences of those
interesting interference phenomena.

\begin{figure}[tbp]
\includegraphics*[width=0.7\linewidth, angle=0]{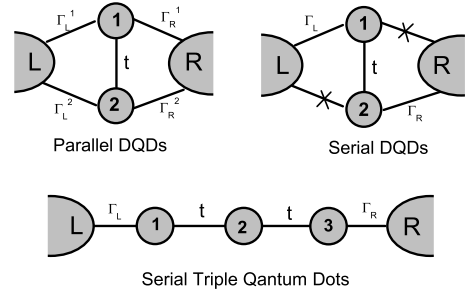}
\caption{Schematic diagrams of coupled multiple quantum dots }
\end{figure}

%Though there are several studies on Fano effects in parallel coupled
%double quantum dots, there is no Fano effect in serially coupled
%double quantum dots due to the lack of interference between a
%discrete state and a set of continuous state. In our present paper,
%our studies will show that Fano resonance can occur in
%photon-assistant transport in serially coupled triple quantum dots
%(STQDs) with the help of ``trapping dark state", which exists in
%atoms with three level structures under certain conditions[17] .
%In a recent article, optically controlled current is studied by
%designing the three-level structure of the system[17] and the effect
%of coherent population trapping is proved to occur in the system.
%However, the system contains only two coupled quantum dots. The
%three-level structure of MQDs involving three coupled quantum dots
%can exhibit more intriguing phenomena as shown in our paper.
In previous theoretical work on MCQDs, the quantum levels, which
depend on the structures of the system, are often assumed as
parameters\cite{18,19}, and the quantum properties of MCQDs can not
be presented quantitatively. Therefore, it is necessary to reveal
the quantum behavior of MCQDs based on more realistic model.
%In this paper, we investigate the formation of a
%$\Lambda$-type three-level structure in MQDs and the related
%quantum effects.
In our approach, we first design DQDs and STQDs (with $\Lambda$-type
three-level structure) using a two-dimensional confining model in
the effective mass frame. Based on the obtained level structure and
with the help of Floquet theory\cite{20}, we study the transport
properties of the DQDS and STQDs.  It is found for both symmetrical
and asymmetrical configurations of coupled DQDs, there is
photon-assisted Fano resonances in parallel DQDs, and a symmetric
resonance in serial DQDs. The electronic transport is {\bf very}
different from that in STQDs with $\Lambda$-type energy level: the
photon-assistant tunneling leads to the symmetric
Breit-Wigner\cite{21} resonance when the system is asymmetric
(between left and right quantum dots). When the system is in a
symmetric (between left and right quantum dots) configuration, the
formation of ``trapping dark state" results in the interesting
asymmetric Fano resonance under resonant condition (photon energy
equals to the energy difference between left/right quantum dot and
the middle dot.) Our studies show that the transport properties are
quite sensitive to the number of quantum dots in the coupled
systems.
%Our studies on STQDs are not a straightforward
%extension of that for double quantum dot. The unique structure of
%STQDs leads to their interesting transport properties.

The organization of the paper is as follows. In section II we
describe the model and calculate the electronic structure of the
MCQDs. In section III we present the photon-assistant transport
properties of the system and discuss the results. A brief summary is
given at the end of the paper.
%The paper is
%summarized in section IV.

\section{Model and Electronic structure}

\begin{figure}[tbp]
\includegraphics*[width=0.8\linewidth, angle=0]{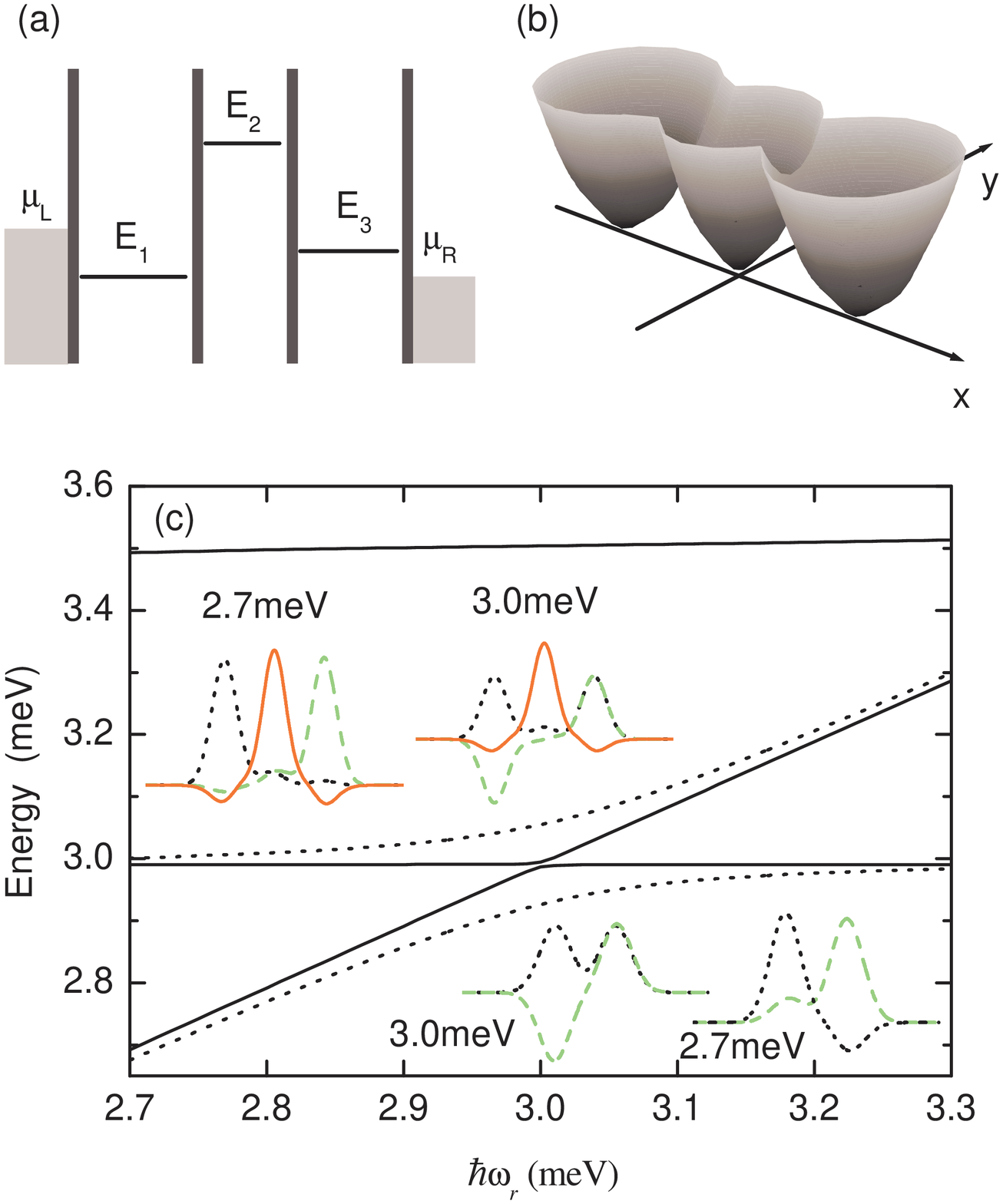}
 \caption{\small (a) Expected $\Lambda$-type three-level scheme for STQDs.
 (b) Confining potential model of STQDs. (c)The three solid lines are levels for a STQD as functions of
 $\hbar\omega_r$ for STQDs of
$\hbar\omega_l=3.00$ meV, $\hbar\omega_m=3.52$ meV, and $d=77.70\
\mathrm{nm}$. The dotted two lines are levels for DQDs of
$\hbar\omega_l=3.00$ meV, $d=77.70\ \mathrm{nm}$. Insert: The
eigenwavefunctions for STQDs (up) and DQDs (bottom) at
$\hbar\omega_r=3.00\ \mathrm{meV}$ and $2.70\ \mathrm{meV}$
respectively.
%The solid line
%is for state $|2\rangle$. The dashed line is for state $|1\rangle$.
%The dotted line is for state $|0\rangle$.
}
 \label{Fig.2}
\end{figure}
We use a two-dimensional model to describe multiple coupled quantum
dots laterally. Here, we give serially coupled triple quantum dots
as an example. The confining potential is
\begin{eqnarray}
V(x,y)&=&0.5m^*\mathrm{min}\{\omega^2_{lx}(x+d)^2+\omega^2_{ly}y^2,
\omega^2_{mx}x^2\nonumber\\
&{}&+\omega^2_{my}y^2,\omega^2_{rx}(x-d)^2+\omega^2_{ry}y^2\},
\end{eqnarray}
where $d$ is the interdot distance, $m^*$ is the effective mass of
electron, $\omega_{lx(y)}$, $\omega_{mx(y)}$ and $\omega_{rx(y)}$
are confining trap frequencies of the left, middle and right dots in
the $x(y)$ direction respectively. The model Hamiltonian of an
electron in the coupled quantum dots can be written as
\begin{eqnarray}
H=\frac{\mathbf{P}^2}{2m^*}+V(\mathbf{r}),
\end{eqnarray}
where $\vec{\mathbf{r}}=(x,y)$. In our calculation, the material
parameter of GaAs QDs is used as $m^*=0.067m_e$ and the value of $d$
is taken to resemble experimental systems. We use the eigenstates
$\varphi_{i_l},\varphi_{i_m},\varphi_{i_r}$ for each dot as the
basis of the Hilbert space. Considering the nonorthogonality of the
basis states, we obtain the eigenstates of Eq. (2)  by solving the
generalized eigenvalue of the system. Here, we investigate the
levels of the STQDs by varying
$\hbar\omega_{rx}=\hbar\omega_{ry}=\hbar\omega_{r}$ with constant
parameters of
$\hbar\omega_{lx}=\hbar\omega_{ly}=\hbar\omega_{l}=3.00$ meV,
$\hbar\omega_{mx}=\hbar\omega_{my}=\hbar\omega_{m}=3.52$ meV, and
$d=77.70\ \mathrm{nm}$. For such parameters, the tunneling energy
between the left and middle dot is about $51\ \mu$eV. The lowest
three energy levels of STQDs form a $\Lambda$-type structure as
shown in Fig. 2(a) (c). Experimentally, the levels of each dot are
tuned by changing the voltage of the corresponding electrode gate.
Such change of voltage corresponds to the variation of confining
strength of the dot in our model. As the value of $\hbar\omega_r$
increases, the right level increases. When
$\hbar\omega_r=\hbar\omega_l=3.00$ meV, the energies of $|1\rangle$
and $|3\rangle$ anticross. The corresponding eigenwavefunctions are
shown in the insert. It is clear that a pair of delocalized bonding
and antibonding states are formed, while the other state is
localized in the middle quantum dot.
 When $\hbar\omega_r$ is away from 3.00
meV,
% Therefore, the states
%$|0\rangle$, $|1\rangle$, and $|2\rangle$ around
%$\hbar\omega_r=3.00$ meV
 the three levels are nearly energies of the ground states of the
left, middle and right dots respectively, and the eigenstates are
all localized in each quantum dots.
%Weconsidered the three quantum dots to be in the so-called
%$\Lambda$-type three-level structure in which the two lower left and
%right levels $E_0$ and $E_1$ are coupled to a single higher middle
%energy level $E_2$.
The continuous manifolds on the two sides of the Fig. 2(a)
correspond to electronic states with chemical potential $\mu_{L}$
and $\mu_{R}$ and the two side dots are coupled to the leads with
the dots-lead hopping rate $\Gamma_L$ and $\Gamma_R$.

Using the same method, we investigate the levels of the DQDs by
varying $\hbar\omega_{r}$ with constant parameters of
$\hbar\omega_{l}=3.00\ \mathrm{meV}$, and $d=77.70\ \mathrm{nm}$.
The lowest two energy levels and the corresponding
eigenwavefunctions with different values of $\hbar\omega_{r}$ of
DQDs as shown in Fig.2(c). Similarly, when
$\hbar\omega_r=\hbar\omega_l=3.00$ meV, the energies of $|1\rangle$
and $|2\rangle$ anticross and a pair of delocalized bonding and
antibonding states are formed. When $\hbar\omega_r$ is away from
3.00 meV, the two levels are nearly energies of the ground states of
the left and right dots respectively, and the eigenstates are
localized in each quantum dots.

\section{Photon-assistant transport properties}
\subsection{Formulism for photon-assistant transport in coupled QDs system}
We first construct our formulism based on parallelly coupled DQDs
(see Fig.1 ) with an external time-varying field. This system can be
described by the following Hamiltonian:
\begin{equation}
H=\sum_{\alpha=L,R}H_{\alpha}+H_{D}+H_{T},
\end{equation}
with
\begin{equation}
H_{\alpha}=\sum_{k}\epsilon_{\alpha,k}a_{\alpha,k}^{\dag}a_{\alpha,k},
\end{equation}
\begin{equation}
H_{D}=\sum_{i=1,2}E_{i}(t)
 d_{i}^{\dag}d_{i}
 -(t_{c}d_{1}^{\dag}d_{2}+H.c.),
\end{equation}
\begin{equation}
H_{T}=\sum_{\alpha,k,i=1,2} t_{\alpha i}
 d_{i}^{\dag}a_{\alpha,k}+H.c..
\end{equation}
$H_{\alpha}$ ($\alpha=L,R$) describes the left and right normal
metal leads. $H_{D}$ models the parallel-coupled DQD where
$d_{i}^{\dag}$ ($d_{i}$) represents the creation (annihilation)
operator of the electron with energy $E_{i}$ in the dot $i$
($i=1,2$). $t_{c}$ denotes the interdot coupling strength, which can
be obtained from the calculations of the electronic structure. Under
the adiabatic approximation, the external ac electric field can be
reflected in the single-electron energies which can be separated
into two parts as $E_{1}(t)=E_{1}+\Delta_{0}(t)$ and
$E_{2}(t)=E_{2}-\Delta_{0}(t)$ for the central conductor. $E_{i}$ is
the time-independent single-electron energies without the ac field,
and $\Delta_{0}(t)$ is a time-dependent part from the ac field,
which can be written as $\Delta_{0}(t)=edA\cos\Omega t$. $H_{T}$
represents the tunneling coupling between the DQD and leads where
$t_{\alpha i}$ is the hopping strength between the $i$th QD and the
$\alpha$ lead. To capture the essential physics of photon-assisted
Fano resonance, we consider the simplest case with noninteracting
electrons in two single-level QDs.

The current $I_{\alpha}(t)$ from the $\alpha$ lead to the central
region can be calculated from standard NGF techniques, and can be
expressed in terms of the dot's Green function
as\cite{Wingreen1,Wingreen2,Sun}
\begin{equation}
I_{\alpha}(t)=\frac{2e}{\hbar}\mathrm{Re}\int dt'
 \mathrm{Tr}\{[\mathbf{G}^{r}(t,t')\mathbf{\Sigma}^{<}_{\alpha}(t',t)
 +\mathbf{G}^{<}(t,t')\mathbf{\Sigma}^{a}_{\alpha}(t',t)]\},
\end{equation}
Here, the Green's function $\mathbf{G}^{r,<}$ and the self-energy
$\mathbf{\Sigma}^{a,<}$ are all two-dimensional matrices for the
DQD system. The bold-faced letters are used to denote matrices.
The retarded and lesser Green functions are defined as
$\mathbf{G}^{r}(t,t')=-i\theta(t-t')\langle\{\Psi(t),\Psi^{\dag}(t')\}\rangle$
and $\mathbf{G}^{<}(t,t')= i\langle\Psi^{\dag}(t')\Psi(t)\rangle$,
respectively, with the operator
$\Psi^{\dag}=(d_{1}^{\dag},d_{2}^{\dag})$.
%In order to obtain the
%expression of the current, we have to solve the Green's functions.
%The retarded Green's functions can be calculated by using Dyson
%equation
%\begin{equation}
%\mathbf{G}^{r}(t,t')=\mathbf{g}^{r}(t,t')
% +\int dt_{1}\int dt_{2}
% \mathbf{G}^{r}(t,t_{1})\mathbf{\Sigma}^{r}(t_{1},t_{2})\mathbf{g}^{r}(t_{2},t'),
%\end{equation}
%where $\mathbf{g}^{r}(t,t')$ is the retarded Green's function for
%the uncoupled DQD without the coupling to the leads.

The total self-energy is
$\mathbf{\Sigma}^{r}=\sum_{\alpha}\mathbf{\Sigma}^{r}_{\alpha}+\mathbf{\Sigma}^{r}_{c}$,
in which $\mathbf{\Sigma}^{r}_{\alpha}$ and
$\mathbf{\Sigma}^{r}_{c}$ are caused by the $\alpha$ lead and the
interdot couplings, respectively. Under the wide-band approximation,
the retarded self-energy caused by the $\alpha$ lead is defined as
\begin{eqnarray}
\mathbf{\Sigma}^{r}_{\alpha}(t,t')
 = -\frac{i}{2}\delta(t-t') \left(\begin{array}{cc}
 \Gamma^{\alpha}_{1}
  & \sqrt{\Gamma^{\alpha}_{1}\Gamma^{\alpha}_{2}}\\
 \sqrt{\Gamma^{\alpha}_{1}\Gamma^{\alpha}_{2}}
  & \Gamma^{\alpha}_{2}
\end{array}\right),
\end{eqnarray}
where $\Gamma^{\alpha}_{i}$ is the linewidth function defined by
$\Gamma^{\alpha}_{i}=2\pi\rho_{\alpha}t^{*}_{\alpha i}t_{\alpha
i}$ with $\rho_{\alpha}$ being the density of states of the
corresponding lead, describing the coupling between the $i$th QD
and the $\alpha$ lead. The advanced self-energy can be obtained
from
$\mathbf{\Sigma}^{a}_{\alpha}(t,t')=(\mathbf{\Sigma}^{r}_{\alpha}(t,t'))^{\dag}$.
The self-energy caused by the interdot coupling is
\begin{eqnarray}
&\mathbf{\Sigma}^{r}_{c}(t,t')
 &= \delta(t-t') \left(\begin{array}{cc}
 0
  & -t_{c}\\
 -t_{c}
  & 0
\end{array}\right).
\end{eqnarray}
%Next, we need to solve lesser Green's functions which is related
%to the retarded Green's function through the Keldysh equation
%\begin{equation}
%\mathbf{G}^{<}(t,t')=\int dt_{1}\int dt_{2}
% \mathbf{G}^{r}(t,t_{1})\mathbf{\Sigma}^{<}(t_{1},t_{2})\mathbf{G}^{a}(t_{2},t').
%\end{equation}
%where
The lesser self-energy is
$\mathbf{\Sigma}^{<}(t_{1},t_{2})=\sum_{\alpha}\mathbf{\Sigma}^{<}_{\alpha}(t_{1},t_{2})$
and
\begin{eqnarray}
\mathbf{\Sigma}^{<}_{\alpha}(t,t')
 = i \int\frac{d\epsilon}{2\pi} f_{\alpha}(\epsilon) e^{-i\epsilon(t-t')}
 \left(\begin{array}{cc}
 \Gamma^{\alpha}_{1}
  & \sqrt{\Gamma^{\alpha}_{1}\Gamma^{\alpha}_{2}}\\
 \sqrt{\Gamma^{\alpha}_{1}\Gamma^{\alpha}_{2}}
  & \Gamma^{\alpha}_{2}
\end{array}\right).
\end{eqnarray}
where
$f_{\alpha}(\epsilon)=1/(e^{(\epsilon-\mu_{\alpha})/k_{B}T}+1)$
denotes the Fermi distribution function of electrons in the
$\alpha$ lead.
%The advanced Green's functions can be calculated as
%$\mathbf{G}^{a}(t,t')=(\mathbf{G}^{r}(t,t'))^{\dag}$.

The Green's function of uncoupled DQD without the couplings to the
two leads can be easily obtained as
\begin{eqnarray}
&\mathbf{g}^{r}(t,t')
 = -i\theta(t-t') \left(\begin{array}{cc}
 e^{-i\int^{t}_{t'}E_{1}(t_{1})dt_{1}} & 0\\
 0 & e^{-i\int^{t}_{t'}E_{2}(t_{1})dt_{1}}
\end{array}\right).
\end{eqnarray}
%Notice that the Green's function depends on two time variables,
%not the time difference. Therefore, one should take the Fourier
%expansion of the Green's function. Then, the Fourier
%transformation of the Green's function is
%\begin{equation}
%\mathbf{g}^{r}_{mn}(\epsilon)=\left(\begin{array}{cc}
% \sum_{l}\frac{J_{l-m}(\alpha)J_{l-n}(\alpha)}{\epsilon-\varepsilon_{1}-l\omega+i0^{+}} & 0\\
% 0 & \sum_{l}\frac{J_{m-l}(\alpha)J_{n-l}(\alpha)}{\epsilon-\varepsilon_{2}-l\omega+i0^{+}}
%\end{array}\right),
%\end{equation}
%where $\alpha=\Delta_{0}/\omega$ and $J_{n}(x)$ is the $n$th
%Bessel function. Here, $\omega$ is the frequency of the ac field.

After taking the Fourier transformations, the Dyson equation and
Keldysh equation become
\begin{equation}
\mathbf{G}^{r}_{mn}(\epsilon)=\mathbf{g}^{r}_{mn}(\epsilon)
 +\sum_{l}\mathbf{G}^{r}_{ml}(\epsilon)\mathbf{\Sigma}^{r}_{ll}(\epsilon)\mathbf{g}^{r}_{ln}(\epsilon),
\end{equation}
and
\begin{equation}
\mathbf{G}^{<}_{mn}(\epsilon)=\sum_{l}
 \mathbf{G}^{r}_{ml}(\epsilon)\mathbf{\Sigma}^{<}_{ll}(\epsilon)\mathbf{G}^{a}_{ln}(\epsilon),
\end{equation}
respectively. With these Green's functions, the time-dependent
current can be expressed as
\begin{equation}
I_{\alpha}(t)=\frac{2e}{h}\mathrm{Re}\sum_{l}e^{i l\Omega t}
 \{\int d\epsilon
 \mathrm{Tr}[\mathbf{G}^{r}(\epsilon)\mathbf{\Sigma}^{<}_{\alpha}(\epsilon)
 +\mathbf{G}^{<}(\epsilon)\mathbf{\Sigma}^{a}_{\alpha}(\epsilon)]_{l0}\}.
\end{equation}
Then, the average current is
\begin{equation}
I=\langle I_{\alpha}(t)\rangle=\frac{2e}{h}\mathrm{Re}
 \{\int d\epsilon
 \mathrm{Tr}[\mathbf{G}^{r}(\epsilon)\mathbf{\Sigma}^{<}_{\alpha}(\epsilon)
 +\mathbf{G}^{<}(\epsilon)\mathbf{\Sigma}^{a}_{\alpha}(\epsilon)]_{00}\}.
\end{equation}

Here we give another formulism suitable for the serially coupled QDs
system. Making use of the Floquet decomposition, the physical
picture is clearer in this formulism. As seen in Fig. 1, we should
have the serially coupled DQD, if we set
$\Gamma_{R}^1=\Gamma_{L}^2=0$. In this case, the self-energies are
simplified. It can be shown that
\begin{equation}
I=\frac{e}{h}\sum_{k=-\infty}^{+\infty}\int
d\varepsilon\{T_{LR}^{(k)}(\varepsilon)f_{R}(\varepsilon)-T_{RL}^{(k)}(\varepsilon)f_{L}(\varepsilon)\},
\end{equation}
where
%\begin{equation}
%T_{LR}^{(k)}(\varepsilon)=\Gamma_{L}(\varepsilon+k\hbar\Omega)\Gamma_{R}(\varepsilon)|\langle1|G^{(k)}(\varepsilon)|N\rangle|^{2}
%\end{equation}
%\begin{equation}
%T_{RL}^{(k)}(\varepsilon)=\Gamma_{R}(\varepsilon+k\hbar\Omega)\Gamma_{L}(\varepsilon)|\langle
%N|G^{(k)}(\varepsilon)|1\rangle|^{2}
%\end{equation}
\begin{equation}
T_{LR}^{(k)}(\varepsilon)=\Gamma_{L}\Gamma_{R}|\langle1|G^{(k)}(\varepsilon)|N\rangle|^{2}
\end{equation}
\begin{equation}
T_{RL}^{(k)}(\varepsilon)=\Gamma_{R}\Gamma_{L}|\langle
N|G^{(k)}(\varepsilon)|1\rangle|^{2}
\end{equation}
(N=2) denote the transmission probabilities for electrons
%with initial
%energy $\varepsilon$ and final energy $\varepsilon+k\hbar\Omega$
from the right lead and from the left lead, respectively.
$f_{\ell}(\varepsilon)=(1+\mathrm{exp}[(\varepsilon-\mu_{\ell})/k_{B}T])^{-1}$
denotes the Fermi function and
\begin{equation}
G^{(k)}(\varepsilon)=\sum_{\beta,k'}\frac{|u_{\beta,k'+k}\rangle\langle
u_{\beta,k'}^{+}|}
{\varepsilon-(\epsilon_{\beta}+k'\hbar\Omega-i\gamma_{\beta})}
\end{equation}
is the Fourier coefficients of the retarded Green function, where
$|u_{\beta,k}\rangle$ are the Fourier coefficients of the Floquet
state $|u_{\beta}(t)\rangle$, $\epsilon_\beta$, $\gamma_\beta$
($\beta=1,2$) are the real and imaginary parts of quasi energies
respectively. This formulism is valid for general N serially coulped
QDs system. \cite{20}

\subsection{Transport properties of DQDs}

\begin{figure}[tbp]
\includegraphics*[width=0.76\linewidth, angle=0]{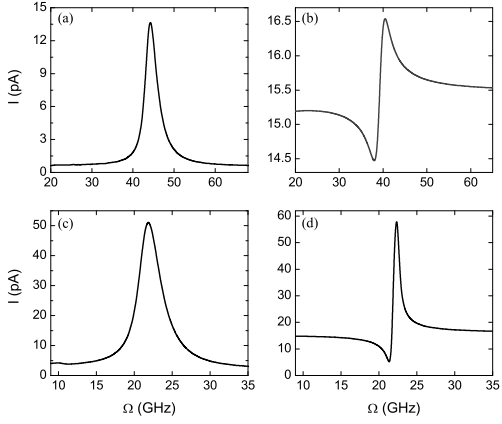}
 \caption{\small (a) Average current $I$ of the serial DQDs.
 (b) Average current $I$ of the parallel DQDs where
 $\hbar\omega_r=3.10$ meV, $k_BT=0$. (c) (d) are the same as (a) (b)
 respectively except that $\hbar\omega_r=3.00\ \mathrm{meV}$. }
\end{figure}

In our numerical calculations, we assume $k_BT=0$ and set the energy
independent dots-lead hopping rate $\Gamma_L=\Gamma_R=9\
\mu\mathrm{eV}$, the applied voltage $\mu_L-\mu_R=15\
\mu\mathrm{V}$, and the ac field magnitude $A=1.54\ \mathrm{V/cm}$.
First we study the transport properties of an asymmetric system by
applying different confining potentials to the left and right dots.
The average current-frequency curves of serially and parallel DQDs
with $\hbar\omega_r=3.10\ \mathrm{meV}$ are shown in Fig.3 (a) and
(b) respectively. We find that the current curve has a symmetric
Breit-Wigner line shape when the two dots are serial, and the
current has an asymmetric Fano line shape when the dots are
parallel. Here the Breit-Wigner line shape is a consequence of
photon-assistant resonant, i.e.,
$\hbar\Omega=\sqrt{(E_{1}-E_{2})^2+4t_c^2}$.
 The new photon-assistant Fano lineshape comes from the interference between photon-modified
bound-antibound channels. Then we apply the same confining
potentials to the two dots and research the electron transport of a
symmetric system. The results with $\hbar\omega_r=3.00\
\mathrm{meV}$ are displayed in Fig.3 (c) and (d). Similarly, we find
that a Breit-Wigner resonance appears in the serially DQDs and a
Fano resonance occurs in the parallel DQDs. From above discussion,
we conclude that the ac electric field can induce the
photon-assisted Fano resonances for both symmetrical and
asymmetrical parallel configurations of DQDs, but can not induce
Fano resonance in the serially DQDs, whether the system is symmetric
or not, which means that the energy level of DQDs does not have
remarkable impact on the transport properties.

\subsection{Transport properties of STQDs}
As  one observes in the last section that there is no Fano effect in
a serially coupled DQD due the lack of apparent interference
channels. One may expect that there is also no Fano resonance in a
STQD. However, our studies shows the transport in STQD is quite
different from that in a DQD, and we will give a quite different
story in this section.

Based on the results of electronic structure in section II, we study
the transport properties through the system with $\Lambda$-type
three-level structure ( Fig.2 ) under the action of an ac driving
field with a frequency $\Omega$.
%The effective
%Hamiltonian of the dots in the basis corresponding to the three
%levels is
%\[
%  \left[\begin{array}{ccc}
%E_l-x_ledE_{ac}(t) & -T_{lm} & 0 \\
%-T_{lm} & E_m-x_medE_{ac}(t) & -T_{rm} \\
%0 & -T_{rm} & E_r-x_redE_{ac}(t)
%\end{array}\right]
%\]
%where $E_{l(m,r)}$ is the on-site energy of the left(middle,right)
%dot, $T_{lm(rm)}$ is the tunneling energy between the left(right)
%and middle dot, $x_{l(m,r)}$ denotes the scaled position of each
%dot, $e$ is the electron charge, $d$ is the interdot distance, and
%$E_{ac}(t)=A\cos(\Omega t)$ is the ac electric field.

%The transport properties of this time-dependent system can be
%studied with the help of Floquet theorem. It can be shown that the
%average current of this system can be written as\cite{20}

In our numerical calculations, we set the applied voltage
$\mu_L-\mu_R=120\ \mu\mathrm{V}$, and the other parameters we have
chosen are the same as the ones for DQDs.

%set the energy independent dots-lead hopping rate
%$\Gamma_L=\Gamma_R=9\ \mu\mathrm{eV}$, and assume that $k_BT=0$,
%then the Fermi functions in the expression for the average current
%(3) become step functions. Other parameters we have chosen are the
%applied voltage $\mu_L-\mu_R=120\ \mu\mathrm{V}$, and the ac field
%magnitude $A=1.54\ \mathrm{V/cm}$.

\begin{center}
\begin{figure*}[tbp]
\includegraphics*[width=0.8\linewidth, angle=-0]{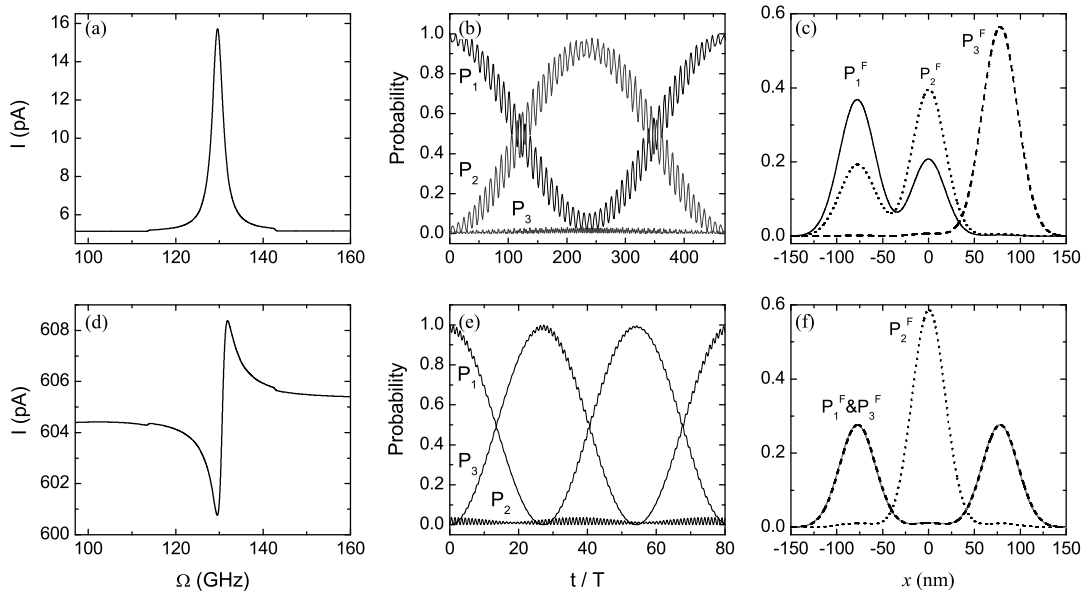}
 \caption{\small (a) Average current $I$ as a function of the driving frequency
 $\Omega$.
  (b) Time-dependent occupation probabilities for
electron in left, middle, right dots. (c) Time average of electron
probability distributions of the Floquet states at $\Omega=129.48\
\mathrm{GHz}$ where
 $\hbar\omega_r=3.10$ meV, $k_BT=0$. (d) (e) (f) are the same as (a) (b)
 (c) respectively
 except that $\Omega=130.57\ \mathrm{GHz}$ and $\hbar\omega_r=3.00\ \mathrm{meV}$. }
 \label{Fig.4}
\end{figure*}
\end{center}

 We first study the transport properties of an asymmetric system by
applying different confining potentials to the left and right dots.
The average current $I$ as a function of the driving frequency
$\Omega$ with $\hbar\omega_r=3.10\ \mathrm{meV}$ is presented in
Fig.4 (a). We find the current curve has a symmetric Breit-Wigner
line shape around $\Omega=129.48\ \mathrm{GHz}$, suggesting that
there is a resonance for electrons in the system in this case
($\hbar\Omega=E_2-E_1$). The time evolution of the probabilities for
an electron in left, middle, right dot are shown in Fig.4 (b). Here
we have performed our calculation in a closed system (i.e.,without
interaction with the leads) and have used the initial condition
$P_L=1$ and the resonant condition $\Omega=129.48\ \mathrm{GHz}$.
Fig. 4(c) shows the time average of the probability distribution of
Floquet states, $P^F_{\beta}=\frac{1}{T}\int^T_0 |u_{\beta}(t)|^2
dt, \beta=1,2,3$. It is clear that there is a photon assistant
mixing between left and middle dot. This photon assistant charge
transfer leads to the occurrence of the current resonance
phenomenon.

Then we consider the case that the system is symmetric by applying
the same confining potential to the two side dots. The average
current $I$ as a function of the driving frequency $\Omega$ with
$\hbar\omega_r=3.00\ \mathrm{meV}$ is displayed in Fig.4 (d). We
find that the current curve has an asymmetric Fano line shape around
$\Omega=130.57\ \mathrm{GHz}$ and the current amplitude is much
larger than that in the asymmetric case. In order to understand the
intriguing phenomenon, we did similar calculation of the time
evolution of the occupation probabilities for an electron in left,
middle, right dot and the time average of probability distribution
of Floquet states. As shown in Fig.4 (e) and (f), we find that two
delocalized Floquet states ( which is related to  bonding and
antibonding states ) are formed and another Floquet state  is still
localized in the middle dot in this case, indicating that the middle
dot mediates the super-exchange interaction between the left and
right dots. The states $|1\rangle,|3\rangle$ behave like ``trapping
dark state" in atomic system.
%Then we apply ac electric field on
%the dots and observe the time-dependent occupation probability of
%electrons on each dot at the frequency $\Omega=820\ \mathrm{GHz}$ in
%a closed system, in which we suppose the electron initially
%localized at the left dot. In Fig.4 (c), the occupation
%probabilities for three states denote that the electron occupy
%mostly the left and right dots, meaning that electrons are in the
%time-dependent trapping state as the case in atomic system. Still
Though state $|2\rangle$ has a very small occupation probability in
the left and right dots,  it plays an important role in the
occurrence of the Fano-type resonance, which can be presented in the
following discussion.

\begin{figure}[tbp]
\includegraphics*[width=0.8\linewidth, angle=-0]{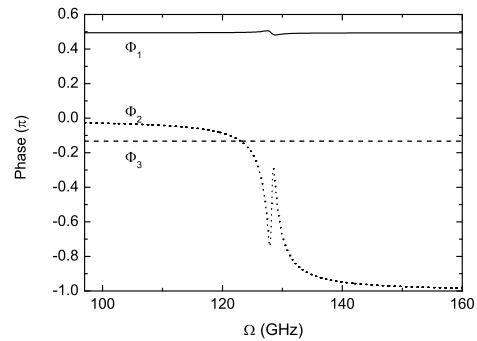}
 \caption{\small  Phases as functions of the driving frequency
 $\Omega$. }
 \label{Fig.5}
\end{figure}

As seen from Eq. (19), three Floquet states form three conducting
channels. Due to the interaction with leads, the three states may
have different width. In the symmetric situation, the bonding and
antibonding Floquet states $|u_1\rangle$, $|u_3\rangle$ have large
occupation probability in left and right dots, thus are strongly
coupled to the leads. They have wide width and can be viewed as
continuous channels. The localized Floquet state $|u_2\rangle$ has
small occupation probability in left and right dots and is weakly
coupled to the leads. It has very narrow width and can be viewed as
resonant discrete channel.  The interference between strongly
coupled channels and weakly coupled channel leads to the observed
Fano resonance\cite{22}. While in the asymmetric situation, the
Floquet state $|u_2\rangle$ is also effectively coupled to the leads
via a photon-assistant mixing between left and middle dot. Thus all
three states form strongly coupled channels, so no Fano effect is
observed. Here we discuss in detail the interference between
different channels. From Eq. (19), one can see that the average
current $I$ is mainly decided by $|\langle
N|G^{(k)}(\varepsilon)|1\rangle|^2$. The interference effects can be
seen from following calculation (here we show term with $k=0$,
$\varepsilon=60\ \mu\mathrm{V}$  as an example)
\begin{equation}
\langle N|G^{(0)}(\varepsilon)|1\rangle=\sum_{\beta,k'}\frac{\langle
N|u_{\beta,k'}\rangle\langle u_{\beta,k'}^{+}|1\rangle}
{\varepsilon-(\epsilon_{\beta}+k'\hbar\Omega-i\gamma_{\beta})}
=\sum_{\beta}A_{\beta} e^{i\phi_{\beta}}.
\end{equation}
%We filter out the items that have minor contributions to the sum
%and select the items that play leading role in the sum, then we
%can write
%\begin{equation}
%\langle N|G^{(0)}(\varepsilon)|1\rangle=\sum_{\alpha}\frac{\langle
%N|u_{\alpha,0}\rangle\langle u_{\alpha,0}^{+}|1\rangle}
%{\varepsilon-(\varepsilon_{\alpha}-i\hbar\gamma_{\alpha})}
%=\sum_{\alpha}A_{\alpha} e^{i\phi_{\alpha}}
%\end{equation}

In Fig.5, we show the phases $\phi_{\beta(\beta=1,2,3)}$ as the
functions of the driving frequency $\Omega$. Here, the phases
$\phi_1$ and $\phi_3$ are phases of two strongly coupled states
$|u_1\rangle$ and $|u_3\rangle$, and $\phi_2$ is the phase of the
weakly coupled state $|u_2\rangle$. From Fig.4, we find that
$\phi_2$ has a phase shift that varies from 0 to $\pi$ as the field
frequency $\Omega$ is moved through the resonance frequency, which
provides resonant channel for electrons to transport the system.
$\phi_1$, $\phi_3$ are nearly constant and can be assumed to be
independent of the driving frequency, which can be considered as
background or nonresonant channels. Therefore, the interference
between the strongly coupled channels and weakly coupled channel
results in the Fano-type resonance. Finally, we would like to point
out that the above picture also explains why there is no Fano
resonance in a serially coupled double dot. As a matter of fact, in
that situation, there are only strongly coupled bonding and
antibonding states. Therefore, there is no apparent Fano effect.

%\section{Conclusion}
In summary, we have studied the transport properties of DQDs and
STQDs under the action of an ac electric field. Using a
two-dimensional confining potential model in the effective mass
frame, the two-level structure and $\Lambda$-type three-level
structure are obtained through solving the generalized eigenvalue of
the system. Based on the level structure and Floquet theory, we
investigate the dependence of the ac current on the electronic
structure of the system. It is found that the two-level structure
does not influence the transport properties of DQDs remarkably. For
both symmetric and asymmetric configurations, there is Fano
resonance in parallel DQDs and no Fano resonance in serial DQDs.
However, it is a different case for STQDs. The $\Lambda$-type
three-level structure has great impact on electron transport: when
the system is asymmetric, the symmetric Breit-Wigner resonance
appears due to phonon assistant tunneling; When the system is
symmetric, the interesting asymmetric Fano resonance occurs under
resonant condition. In this case, quantum interference results in
the formation of trapping dark states. These trapping dark states
are the delocalized bonding and antibonding states which serve as
the continuous channels. Here the middle dot plays a dual role: (1)
It mediates the super-exchange interaction between the left and the
right dots; (2) The localized state within it serves as the resonant
discrete channel. Our work presents the unique quantum interference
features of the multiple quantum dots system and is useful for the
design of novel nanodevices. In short, for transport in a system
with N quantum dots, the situation with N=1, is different from that
with N=2 and the case with N=2 is different from that with N=3. We
do not expect qualitative difference for system with $N>3$.

This work is supported in part by the National Natural Science of
China under No. 10574017, 10774016 and a grant of the China Academy
of Engineering and Physics.

%*Corresponding author, Email-address: duan$\_$suqing@iapcm.ac.cn

\end{document}